\documentstyle[12pt,epsfig]{article}
\begin{document}

\begin{flushright}
{\large
INR-0998/98\\
February 1998 }
\end{flushright}

\vspace{1.5cm}

\begin{center}
{\large \bf On search for a new light gauge boson from 
$\pi^{0}(\eta)\rightarrow\gamma + X$ decays in neutrino experiments} 
\end{center}
\vspace{0.5cm}

\begin{center}  
S.N.~Gninenko\footnote{E-mail address:
 Sergei.Gninenko\char 64 cern.ch} and N.V.~Krasnikov\footnote{E-mail address:
nkrasnik\char 64 vxcern.cern.ch}\\
{\it Institute for Nuclear Research of the Russian Academy of Sciences, Moscow 117312}
\end{center}

\begin{abstract}
 It is shown that a new light gauge boson $X$ which might be produced in the 
decays of pseudoscalar mesons $\pi^{0}(\eta)\rightarrow\gamma + X$
could be effectively searched for in  neutrino experiments 
via the Primakoff effect, in the process of   
$X + Z\rightarrow \pi^{0}(\eta) + Z$
conversion in the external Coulomb field of a nucleus.\ An estimate 
of the $X\rightarrow \pi^{0}$ conversion rate for the NOMAD neutrino detector at CERN
is given.
\end{abstract}

\section{Introduction}
New neutral gauge bosons $X$ are predicted by many models addressing the 
physics beyond the  Standard Model (SM) such as  GUTs \cite{1}, 
supersymmetric \cite{2}, superstring
models \cite{3} and models including a new long-range interaction, i.e. the fifth force, \cite{4}.\
The predictions for the  mass of the  $X$ boson are not very firm and it 
could be light enough ($M_{X}\ll M_{Z}$) for searches  at low 
energies.\

The detailed study of the possible manifestations of light gauge boson 
 was performed in \cite{5,6}.\ It was shown that  
if the mass $M_{X}$ is of the order of the pion mass, an effective search 
could be performed for this new vector boson in  
 the radiative decays of neutral pseudoscalar mesons 
 $P\rightarrow\gamma + X$, where $P = \pi^{0},\eta$, or $\eta^{\prime}$.\

From the analysis of the 
data from earlier experiments, 
 constraints on branching ratios for the decay of  $P\rightarrow\gamma + X$ 
range from  $10^{-7}$ to $10^{-3}$ depending on whether $X$ interacts with 
both quarks and leptons or only with quarks.\ In the first case 
$X$ is a short lived particle decaying mainly to $e^{+}e^{-}$ or $\nu \overline{\nu}$ pairs, while in the second case 
$X$ should be a  relatively long lived particle ($\tau_{X} \geq 10^{-6} sec$)
 \cite{6}.   

 Direct searches for a signal from 
$\pi^{0} \rightarrow\gamma + X$ decay  have been performed in a
 few experiments.\
   The best experimental limit on the decay
$\pi^{0}\rightarrow\gamma + X$ was obtained 
recently by the Crystal Barrel Collaboration at CERN
using $\it p\overline{p}$
annihilations as a source of pseudoscalar mesons, see \cite{7} and 
references there.\
The branching ratio limit of $(3$ to $0.6)\times 10^{-4}$  ($90\%~ C.L.$) 
has been obtained for $0<M_{X}<125$ MeV.\  
Limit of the same order 
has been obtained for $BR(\eta\rightarrow\gamma+X)$  for the higher $M_{X}$
mass region.\ The results are valid for the case where $X$ is a long lived
particle or the $X$ boson decays preferably into $\nu\overline{\nu}$ pairs.

In this paper we show that a new light relatively long lived gauge boson which might be produced 
in decays of $\pi^{0}(\eta) \rightarrow 
\gamma + X$ could be effectively searched for in neutrino experiments via the 
Primakoff effect, in the process of   
$X Z\rightarrow \pi^{0}(\eta) Z $
conversion in the external Coulomb field of a nucleus.\

The paper is organised as follows. In section 2 we discuss the 
X-boson phenomenology. Section 3 describes the suggested 
method and presents the results of calculations of the cross section for $X \rightarrow 
\pi^{0}$ conversion.\ An estimate of the total cross section for the 
X-boson interaction with the matter is given in section 4.\ In section 5 we consider
 NOMAD neutrino detector at CERN  as example to estimate   
expected $X\rightarrow\pi^{0}$ conversion rate in this experiment.\ Section 6 contains 
concluding remarks and discussion.\

\section{X-boson phenomenology}

As it has been mentioned in the introduction many modern models predict 
an enlargement of the standard $SU(3) \otimes SU(2) \otimes U(1)$ 
gauge group by an extra $U(1)$ factor. At present there are no firm 
theoretical prediction on the mass of new gauge boson so we can't 
exclude the possibility that new gauge boson is rather light with a mass 
$M_{X} \leq O(m_{\pi})$. The possibility of the existence of a 
vector boson, which interacts with the baryon current, was suggested 
by Lee and Yang \cite{8}. If this boson is massless, then the results of 
the experiments on the check of the equivalence principle imply a strong 
constraint on its coupling constant \cite{9} $\alpha_{LY} \leq 10^{-47}$. 
However this constraint is no longer valid for massive gauge boson with 
microscopic Compton wavelength. Astrophysical bound on the X-boson 
coupling is \cite{10} $\alpha_{X} \leq O(10^{-10})$ for $M_{X} \leq 
O(100)$ KeV. 

The interaction Lagrangian of the X-boson with quarks and leptons has the 
form \cite{6}
\begin{equation}
L_{X,ql} = g_X[\sum_{q}(g_{Vq}\bar{q}\gamma_{\mu}q + g_{Aq}\bar{q}
\gamma_{\mu}\gamma_{5}q) + \sum_{l}(g_{Vl}\bar{l}\gamma_{\mu}l + 
g_{Al}\bar{l}\gamma_{\mu}\gamma_{5}l)]X^{\mu} , 
\end{equation}
where $(g_{Vq}, g_{Aq}, g_{Vl}, g_{Al}) \sim O(1)$. By C-parity arguments, 
 one can find that a coupling of X-boson to the axial quark current 
gives a negligible small contribution to the $\pi^{0} \rightarrow 
\gamma + X$ decay width. The $\pi^{o} \rightarrow \gamma X$ decay width is 
determined by the formula \cite{6}
\begin{equation}
\Gamma(\pi^{0} \rightarrow \gamma + X) = \frac{2\alpha \alpha_{X}}{(4\pi)^3}
(2g_{Vu}+g_{Vd})^2\frac{m^3_{\pi}}{f_{\pi}^2}(1-\frac{M_{X}^2}{m^2_{\pi}})^3,
\end{equation}      
where $\alpha_{X} = \frac{g^2_{X}}{4\pi}$, $f_{\pi} = 93 MeV$ and $M_{X}$ 
is the mass of the $X$-boson. For the $Br(\pi^{0} \rightarrow \gamma + X)$ 
we have \cite{6}
\begin{equation}
Br(\pi^{0} \rightarrow \gamma + X) = 2 \frac{\alpha_{X}}{\alpha}
(2g_{Vu}+g_{Vd})^2 (1-\frac{M^2_{X}}{m^2_{\pi}})^3
\end{equation}

For the case when $X$-boson interacts with both quarks and leptons combined 
bounds 
from anomalous magnetic moment of leptons, elastic 
$\nu_{e}e \rightarrow \nu_{e}e$ scattering and  beam dump 
experiments lead \cite{6} to 
\begin{equation}
\alpha_{X} \leq O(10^{-9}),
\end{equation}
or to 
\begin{equation}
Br(\pi^{0} \rightarrow \gamma + X) \leq O(10^{-6})
\end{equation}
The $X$-boson lifetime for $M_{X} \geq 1 MeV$ and $\alpha_{X} = O(10^{-9})$ 
is less than $\tau(X) \leq O(10^{-12}) $ sec. So, $X$-boson is short lived 
particles and such scenario is not interesting for us.\ 

In this letter we 
consider the case where the $X$  is a ``leptophobic'' boson which 
interacts only with quarks.\ Consider the life time of the $X$-boson for such
scenario.\  
Experimental limits for the $Br(\pi^{0}(\eta) \rightarrow \gamma + X)$
\cite{7} lead to the bound on coupling of $X$ to quarks of 
\begin{equation}
\alpha_{X} \leq 10^{-6}
\end{equation}
\ For $M_{X} < 2m_{e}$ the $X$-boson decay width into 3 photons 
via quark loop is proportional to 
\begin{equation}
\Gamma(X \rightarrow 3 \gamma) \sim \alpha^{3}\alpha_{X}
(\frac{M_{X}}{m_{q,eff}})^8  M_{X}, 
\end{equation}
where $m_{q,eff} \approx 300 MeV$ is an effective quark mass. Using Eq.(6) 
we find that $\tau_{X}(X \rightarrow 3 \gamma) \geq 
O(10)$ sec. The decay width of the $X$-boson into $\nu \bar{\nu}$ via quark loop 
is proportional to 
\begin{equation}
\Gamma(X \rightarrow \nu\bar{\nu}) \sim \alpha_{X}G^2_FM^5_X
(\frac{M_X}{m_{q,eff}})^4
\end{equation}
For  $M_{X} \leq 100$ MeV we find that 
$\tau_{X}(X \rightarrow \nu \bar{\nu}) \geq O(10^{-4})$ sec. For $M_{X} > 2m_{e}$ 
the $X$-boson decays mainly into $e^{+}e^{-}$ through the quark loop and 
its decay width  for $M^{2}_{X} \ll m^{2}_{q,eff}$ is
\begin{equation} 
\Gamma(X \rightarrow e^{+}e^{-}) \simeq \frac{1}{12\pi}(\frac{\alpha}{15\pi})^2
[3(\frac{2}{3}g_{Vu}-\frac{1}{3}g_{Vd}-\frac{1}{3}g_{Vs})^{2}]^{2} 
\alpha_{X} M_{X} (\frac{M_{X}}{m_{q,eff}})^4
\end{equation}
For  $M_{X} \leq 100$ MeV and $[3(\frac{2}{3}g_{Vu}-\frac{1}{3}g_{Vd}-\frac{1}{3}g_{Vs})^{2}]^{2} = O(1)$ we find that 
$\tau_{X}(X \rightarrow e^{+}e^{-}) \geq O(10^{-6})$ sec.\  So, we conclude 
that for the case when the X-boson interacts only with quarks it is 
long lived particle at least for $M_{X} \leq 100$ MeV and for $X$ boson energy
$E_{X} > 10$ GeV it travels 
the distance more than 10 km before its decay which is much larger than 
the typical decay length ($\simeq$1 km)  in the current neutrino experiments. 

\section{Method of search and cross section for $X\rightarrow \pi^{0}$ conversion }

Let us consider for simplicity the decay $\pi^{0}\rightarrow\gamma+X$.\ The 
consideration for decay of $\eta$  is similar.\
 If the decay $\pi^{0}\rightarrow\gamma+X$ exists, one expects a flux
 of high energy $X$ bosons
 from a neutrino target, since $\pi^{0}$ are abundantly produced
in the forward direction  by high energy ( a few hundred  GeV) protons  
 in a neutrino  target.\ If $X$ is a long lived particle, 
this flux  would penetrate the downstream shielding 
without significant attenuation (see section 4) and would be 
observed in a neutrino detector via the Primakoff effect, namely in the 
 conversion process $X\rightarrow \pi^{0}$ in the external Coulomb field of a 
nucleus (see Figure 1).\ 

 Because the cross section for $X \rightarrow \pi^{0}$ conversion is 
proportional to Z$^{2}$, preferable  search for such events is  in  
high-Z  detectors.\
The experimental signature of $X \rightarrow \pi^{0}$ conversion is a  
single  high energy
$\pi^{0}$ decaying into two photons  which results in 
isolated electromagnetic showers in the detector.\ 
  The occurrence of $X\rightarrow  \pi^{0}$ conversion  would appear as an 
excess of
neutrino-like interactions with pure electromagnetic final 
states above those expected from  SM predictions.\ Note that 
$X \rightarrow \eta$ conversion could also be identified 
through charged decay modes of $\eta$ mesons in the final state.\  

 The  energy spectrum of $X$-bosons at neutrino detector is
expected to be harder than that of neutrinos, since there is no suppression
of the high energy part due to life time of $\pi^{0}(\eta)$,
 as it is in case of
  charged pions those decays in flight are the main source of
neutrino.\ Thus, one could expect that 
the highest energy bins have better sensitivity to $X\rightarrow \pi^{0}$ conversion
signal.\

 \begin{figure}
   \mbox{\epsfig{file=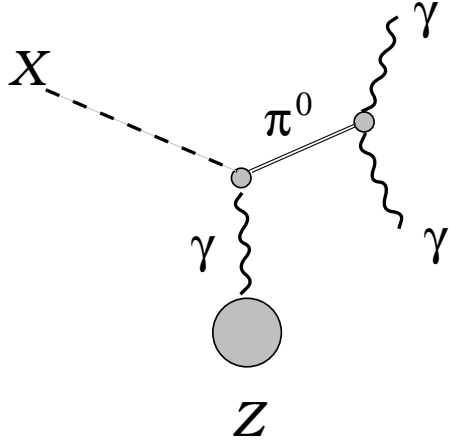,height=60mm}}
    \centering
  \caption{\em Feynman diagram for production of $\pi^{0}$ by Primakoff 
effect.}
  \label{mu1}
\end{figure}

The cross section for $X \rightarrow \pi^{0}$ conversion via the Primakoff 
mechanism is given by the minimal modification of the corresponding formula 
for the axion photoproduction \cite{11}.\ In the lab system, we find
 
\begin{equation}
\frac{d \sigma}{d \Omega}(X + Z \rightarrow \pi^{0} + Z) = 
\frac{1}{2} Br(\pi^{0} \rightarrow X \gamma) \frac{8 \Gamma (\pi^{0} 
\rightarrow \gamma \gamma)}{m_{\pi}^3 \Delta } \alpha Z^2F^2(t)
\frac{P^4 \sin^{2}(\theta)}{t^2} ,
\end{equation}

where $Br(\pi^0 \rightarrow X \gamma) = \frac{\Gamma(\pi^0 \rightarrow 
\gamma + X)}{\Gamma(\pi^0 \rightarrow \gamma\gamma)}$   , 
$P$ is the momentum of the $X$-boson, $\alpha = \frac{1}{137}$, 
$Z^{2}F^{2}(t)$ is the target form factor, $t = -(p_{1} - p_{3})^2$ and  
$\Delta = (1 - \frac{m_{X}^2}{m_{\pi}^2})^3$.\ Since the  minimal momentum square in 
our case $t_{min} = (\frac{m^2_{X} - m^2_{\pi}}{2P})^2$,
 is rather small, atomic form-factors must be used when t is 
small and a nuclear form factor for $ t \geq t_{0} = 
7.39 \cdot m^{2}_{e}$ \cite{11}-\cite{13}.\ It should be noted that in 
comparison with 
 formula (18) of ref.\cite{11} which describes the cross section of axion 
photoproduction ($\gamma + Z \rightarrow a + Z$), a factor of $\frac{1}{2}$ 
arises from  photon 
and X-boson non-identity  and the  factor $\Delta$ in the denominator comes from the kinematics 
of $\pi^{0} \rightarrow \gamma + X$ decay.\ The target form-factor $Z^{2}F^{2}(t)$
consists of three parts.\ At small $t \leq t_{0}$, we
use Thomas-Fermi-Moliere model for the atomic form-factors \cite{13}:
\begin{equation}
Z^2 F^2(t) = G^{el}(t) + G^{inel}(t),
\end{equation}
\begin{equation}
G^{el}(t) = Z^2\frac{a^4t^2}{(1+a^2t)^2},
\end{equation}
\begin{equation}
G^{inel}(t) = Z\frac{a^4_1t^2}{(1 + a^2_1t)^2},
\end{equation}
where $a = \frac{111.7 \cdot Z^{-\frac{1}{3}}}{m_{e}}$ , 
$a_1 =\frac{724.2\cdot Z^{-\frac{2}{3}}}{m_e}$.\ 
For values of $t \geq t_0 $, we use the 
elastic nuclear  form-factor \cite{13}
\begin{equation}
Z^2 F^2(t) = G^{nucl}(t),
\end{equation}
\begin{equation}
G^{nucl}(t) = \frac{Z^2}{(1 + \frac{t}{d})^2}
\end{equation}
where $d = 0.164A^{-\frac{2}{3}} GeV^{2}$ and A is the mass number.\ 
For $t$- values relevant to  our case the differential cross section 
$\frac{d\sigma}{dt}$ can be written in the form
\begin{equation}
\frac{d\sigma}{dt} \simeq \frac{1}{2}Br(\pi^0 \rightarrow X \gamma)
\frac{8\pi \Gamma(\pi^0 \rightarrow \gamma\gamma)}{m^3_{\pi}\Delta}
\cdot\alpha \frac{Z^2F^2(t)}{t},
\end{equation} 
where $ t \simeq 2P^{2}(1 - \cos\theta)$ for heavy nuclei. Using the atomic 
and nuclear form-factors (2-6) we find that the total cross section 
for $t_{0} \geq  t_{min}$ is given by  
\begin{equation}
\sigma(X + Z \rightarrow \pi^0 + Z) =
Br(\pi^0 \rightarrow \gamma + X)\frac{8\pi\Gamma(\pi^0 \rightarrow \gamma 
\gamma)}
{m^3_{\pi}\Delta} \cdot\alpha (G_1 + G_2 + G_3),
\end{equation}
\begin{equation}
G_1 = \frac{Z^2}{2}[ln(\frac{a^2t_0 +1}{a^2t_{min} +1}) + 
\frac{1}{a^2t_0 +1} - \frac{1}{a^2t_{min} + 1}] ,
\end{equation}
\begin{equation}
G_2 = \frac{Z}{2}[ln(\frac{a^2_1t_0 +1}{a^2_1t_{min} +1}) +
\frac{1}{a^2_1t_0 + 1} - \frac{1}{a^2_1 t_{min} + 1}],
\end{equation}
\begin{equation}
G_3\simeq Z^2(ln(\frac{\sqrt{d}}{m_e}) - \frac{3}{2})
\end{equation}

For heavy nuclei and  for $X$ boson masses 
 not too close to the $\pi^0$ mass and for 
$P \leq 100$ GeV, we find that $a_{1}^2t_0 \gg 1$, $a_{1}^2t_{min} \gg 1 $.\ 
The inelastic atomic  form-factor gives a contribution to the total cross section 
of about 1$\%$, which we neglect.\ The total 
cross section then  takes the simple form:
\begin{equation}
\sigma(X + Z \rightarrow \pi^0 + Z) \approx 
Br(\pi^0 \rightarrow X \gamma)\frac{8\pi\Gamma(\pi^0 \rightarrow 
\gamma \gamma)}
{m_{\pi}^3 \Delta}Z^2\cdot \alpha \cdot [ln(\frac{2P\sqrt{d}}{(m^2_{\pi} - 
M^2_X}) - \frac{1}{2}]
\end{equation}
For Pb ($Z = 82$,  $A = 207$), we find  $a = 50.5$MeV$^{-1}$, 
$a_{1} = 75.3 $MeV$^{-1}$, $\sqrt{d} = 68.2$ MeV. Note that the total 
cross section (12)  depends neither on the atomic radius $a$, $a_{1}$ nor on 
the value $t_0$ (the border between the application of atomic and nuclear 
form-factors). 

Numerically for $\Gamma(\pi^0 \rightarrow \gamma \gamma) = 7.7$ eV and for 
$X$ boson masses $M_X < 100$ MeV, 
we find that the total cross section of $X\rightarrow \pi^{0}$ conversion on lead 
depends rather weakly on the incoming $X$ boson momentum  as shown in Figure 2.

The approximation (12) works up to $M_X \leq 125$ MeV.\ For $M_X$ close 
to $m_{\pi}$ it is necessary to use  formulae (8-11).\ The accuracy of the
form-factor calculations is estimated to be better than 5$\%$, \cite{13}; so
the accuracy of our formula (12) is of the same order of magnitude. 

\begin{figure}[hbt]
\begin{center}
   \mbox{\epsfig{file=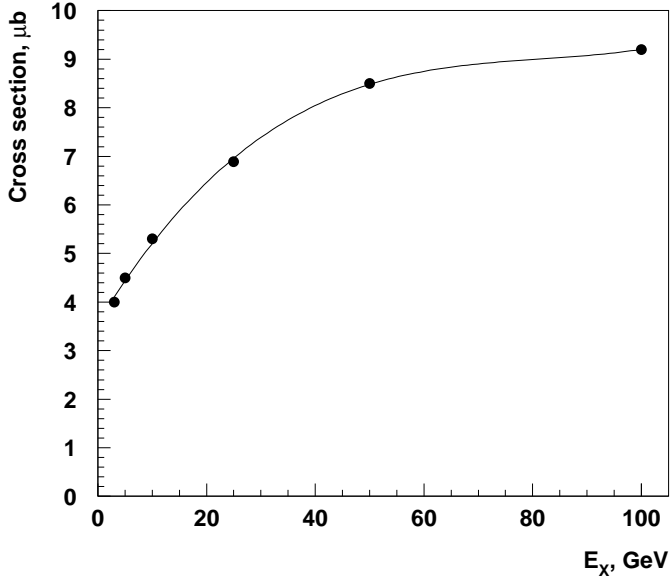,height=90mm}}
  \caption{\em  Cross section for $X \rightarrow \pi^{0}$ conversion on 
    lead versus $X$ boson energy calculated for $M_{X} = 10~ MeV$ and
$Br(\pi^{0}\rightarrow\gamma + X) = 1$.\ The curve is a polynomial fit
to the points.}     
\end{center}
  \label{figure 19:}
\end{figure}

\section {Estimate of the total cross section for  $X$ interactions with  
matter}

The total cross section $\sigma_{t}(X + Fe \rightarrow all)$ for $X$-interactions
in the Fe- shielding used in neutrino experiments  can be estimated in the following way.
We consider the particular case where the $X$ boson interaction  
with u- and d-
quarks is proportional to the electromagnetic interaction, namely:
\begin{equation}
L_i = g_{X}X^{\mu}[\frac{2}{3} \bar{u} \gamma_{\mu} u - \frac{1}{3} \bar{d}
\gamma_{\mu} d]
\end{equation}
In this case, one can find that the total $X$-proton interaction cross section 
is  
\begin{equation}
\sigma_t(Xp) = \frac{1}{2} Br(\pi^{0} \rightarrow \gamma + X) 
(1 -(\frac{m_X}{m_{\pi}})^2)^{-3} \sigma_t(\gamma p)
\end{equation}
The $X$ boson cross section  on the nuclei in the lowest 
order approximation is proportional to the atomic number  in full 
analogy with the case of neutrino scattering (here we implicitly suppose that 
the cross section on protons is equal to the cross section on the neutrons, 
an approximation valid to within a  factor of 2).

\begin{equation}
\sigma_t(xA) \approx A \sigma_t(X p)
\end{equation}
Numerically for laboratory beam momenta 1 GeV$ \leq P_L \leq 10^3$ GeV,
\cite{14} one has:
\begin{equation}
\sigma_t(\gamma p) = (0.12 - 0.16) mb
\end{equation}
For Iron (A =56) and for $M_{X}$ much smaller than $m_{\pi}$ we 
find that
\begin{equation}
\sigma_t(xA) \approx Br(\pi^{0} \rightarrow \gamma + X)\times 4 mb
\end{equation}
For $Br(\pi^{0} \rightarrow \gamma + X) \leq 10^{-4}$, we find  
\begin{equation}
\sigma_t(XA) \leq 4 \times \cdot 10^{-4} mb
\end{equation}

 This estimate of the total cross section for $X$ interaction with matter shows that 
the assumption that $X$ bosons are penetrating  particles is correct 
since for $Br(\pi^{0} \rightarrow \gamma + X) \leq 10^{-4}$ we find that the $X$ boson mean free path in iron is $\geq$ 300 km, as compared 
with the Fe and earth shielding ( total length $\approx$ 0.4 km) used for
example in the CERN SPS neutrino  beam.  

\section{ Rate of $X\rightarrow \pi^{0}$ conversion in the NOMAD detector.}

In this section we  consider the  NOMAD neutrino detector at CERN\cite{15} as an example
in order to estimate the rate of  $X \rightarrow \pi^{0}$ conversion
in this experiment.\ The neutrino beam  is generated  by 450 GeV protons 
delivered by the SPS to the Be neutrino target.\ 

The expected number of $X\rightarrow\pi^{0}$ events can be 
 calculated using the following relations:
\begin{equation}
N_{X-\pi^{0}}=  [Br(\pi^{0}\rightarrow\gamma + X)]^2 \cdot N_{pot} \cdot
\int f_{0}(M_{X}) \sigma_{0}(M_{X},E_{X}) \varepsilon_{sel} dE_{X} \cdot 
\frac{M_{t}}{S} \cdot \frac{N_{A}}{A}
\end{equation}

where $N_{X-\pi^{0}}$ is the predicted number of $X\rightarrow\pi^{0}$ 
events for the given $Br(\pi^{0}\rightarrow\gamma + X)$,
$\varepsilon_{sel}$ is the detection
efficiency,
$f_{0}(M_{X})$ is the flux of $X$ bosons per
one proton on neutrino target (pot), $N_{pot}$ is the total number of pot's,
 $\sigma_{0}(M_{X},E_{X})$ is the cross section of $X \rightarrow \pi^{0}$ conversion
on target with mass number $A$, calculated for  $Br(\pi^{0}\rightarrow\gamma + X)$=1, and $M_{t}$ and $S$ are fiducial mass and fiducial
 area 
of the detector target, respectively.\ 
 We note that
 $Br(\pi^{0}\rightarrow\gamma + X)$ appears twice 
in the formula 
for $N_{X-\pi^{0}}$, through  the $X$ boson flux from the target, and
through the  Primakoff mechanism.\

The flux of the $X$- bosons through NOMAD detector was calculated with 
a detailed GEANT \cite {16} simulation used to predict the neutrino flux
at the NOMAD detector, and was found to be of the order  $f_{0}(M_{X})\simeq 
0.1 X /2.4 \times 2.4~ m^{2}/pot$ for  $Br(\pi^{0}\rightarrow\gamma + X)$=1
and with average $X$-boson energy $<E_{X}>\simeq 50~GeV$.\ Note that  the flux and average energy 
$<E_{X}>$ are  weakly depended on mass $M_{X}$.\ This results in 
weak dependence of limit on branching ratio versus $M_{X}$.

We will consider  NOMAD preshower detector\cite{15} as a lead target  with  
$M_{t} \approx $1 ton and $S = 2.4 \times 2.4 m^{2}$. Using 
$N_{pot}$ = 10$^{18}$ pot 
, $Br(\pi^{0}\rightarrow\gamma + X)$ = 3$\times$
10$^{-4}$, which corresponds to the  Crystal Barrel upper limit for 
$M_{X}\leq 10 MeV/c^{2}$, $\varepsilon_{sel} = 1$ and value of the cross section 
 $\sigma_{0}(M_{X},E_{X})$ = 4.0 $\mu$b taken for simplicity at
 $E_{X}$= 5 GeV (see Figure 2) one obtains

\begin{equation}
N_{X-\pi^{0}}\simeq  Br(\pi^{0}\rightarrow\gamma + X)^{2} \cdot f_{0} 
\cdot N_{pot} \cdot \sigma_{0} \cdot \varepsilon_{sel} \cdot \frac{M_{t}}{S} \cdot \frac{N_{A}}{A} 
\simeq 2\times 10^{3}~events/10^{18}pot
\end{equation}

The main contribution to the background for the 
$X\rightarrow \pi^{0}$ conversion is expected from 
the neutrino  processes  which have a significant electromagnetic
 component in the final state, e.g. coherent  and  diffractive $\pi^{0}$ production, quasi-elastic $\nu_{e}$ scattering, etc..\ Using the neutrino 
cross sections which are well known, the total 
number of background events $N_{bkgd}$ was estimated to be $\simeq 10$ events.\
Thus, if the decay  $\pi^{0}\rightarrow \gamma + X$ exists
with the branching ration of the order of Crystal Barrel limit its effect would be easy seen at NOMAD detector.\ The limit on $Br(\pi^{0}\rightarrow\gamma + X)$ could be  improved  by a factor  
of the order $\sqrt{N_{X-\pi^{0}}/N_{bkgd}} \simeq 10$.\ The coupling to 
quarks could be constrained to $\alpha_{X} \leq 10^{-7}$.

The similar estimate for $X\rightarrow \eta$ conversion rate shows that a limit
of the order of the Crystal Barrel limit on $Br(\eta\rightarrow \gamma + X)$
could be obtained, however for a longer exposure  time at neutrino beam,
since the production of $\eta$ mesons in the forward direction 
in high energy proton collision with neutrino target  is suppressed 
by a factor of $\simeq 10^{2}$ compare to that of 
$\pi^{0}$'s.

\section {Conclusion}

 We have shown that a new light gauge boson $X$ which might be produced in the 
decays of pseudoscalar mesons $\pi^{0}(\eta)\rightarrow\gamma + X$
could be effectively searched for in  neutrino experiments .\ The $X$'s 
being produced in neutrino target from these decays 
would  penetrate the downstream shielding , and be observed in a neutrino detector
via the Primakoff effect, the process of   
$X + Z\rightarrow \pi^{0}(\eta) + Z$
conversion in the external Coulomb field of a nucleus.\\ 
 Our estimates show, that NOMAD experiment at CERN 
is able to improve current limit on $Br(\pi^{0}\rightarrow \gamma + X)$ at least by 
a factor of 10 for the mass region $M_{X} \leq 10~MeV $ during 
$\simeq$ 1 month of exposure at neutrino beam.\

Note that direct searches of these rare radiative  meson decays
 were performed in the previous experiments \cite{7} by searching for a
peak in inclusive photon spectra from  two-body decay of the tagged 
$\pi^{0}(\eta)$ mesons, where background due to missing of one of the two photon 
from principal decay mode $\pi^{0}(\eta)\rightarrow 2\gamma$ is dominated for the 
low $M_{X}$ mass region.\ Our method is free from this 
disadvantage since  its  sensitivity  is
weakly depended on $M_{X}$.  
Note also that another promising way to search for a new light gauge boson from the 
pseudoscalar meson  decays in neutrino experiments 
 is to analyse the ratio $R = \sigma (\nu NC)/\sigma (\nu CC)$ which might 
result in better sensitivity.\ However, in this case a 
knowledge of the  final state structure in $X$ interactions with a quarks 
is required.\ This subject is now under study.\ 

\vspace{1.0cm}

{\Large \bf Acknowledgements}

\vspace{1.0cm}
We are indebted to John Ellis for useful conversation.\ 
One of the authors (S.N.G.) would like to thank his colleagues from the 
NOMAD collaboration for fruitful discussions and support.\ The support of 
the Institute for Nuclear Research, Moscow is gratefully  acknowledged.

\end{document}